# Aspects of Superdeterminism Made Intuitive

Vitaly Nikolaev, Louis Vervoort[(1)]

*School of Advanced Studies, University of Tyumen, Russia*

[(1)] Corresponding author, l.vervoort@utmn.ru

12.02.2022

**Abstract**. We attempt to make superdeterminism more intuitive, notably by simulating a deterministic model system, a billiard game. In this system an initial 'bang' correlates all events, just as in the superdeterministic universe. We introduce the notions of 'strong' and 'soft' superdeterminism, in order to clarify debates in the literature. Based on the analogy with billiards, we show that superdeterministic correlations may exist as a matter of principle, but be undetectable for all practical purposes. This allows us to counter classical objections to superdeterminism such as the claim that it would be at odds with the scientific method, and with the construction of new theories. Finally, we show that probability theory, as a physical theory, indicates that superdeterminism has a greater explanatory power than its competitors: it can coherently answer questions for which other positions remain powerless.

**1. Introduction**.

Superdeterminism (SD) has been discussed as a theoretical loophole for Bell's no-go theorem since many years (early discussions can be found in [1-3]). Searching for escape routes from Bell's no-go verdict seems important: the theorem does nothing less than tolling the bell of death for theories that complete quantum mechanics in a relativistic (i.e. local) way. Until recently SD had few adepts: many believe it has perplexing implications, such as conspiratorial correlations between events, precluding the normal practice of science. Yet, in the background of the still lacking intuitive understanding of the foundations of quantum mechanics, and of the lasting problem to unify quantum field theories and general relativity, SD has recently attracted increased interest [4-16] – and increased criticism [17-21]. In this article, we wish to contribute to clarifying some aspects of SD; by the same token we hope to rebut some of its criticisms.

The most frequent objections to SD are (these criticisms are intimately related but differentiated in the literature): C1) it would preclude doing science because it posits a statistical dependence between observed properties and the choice of test conditions (e.g. [1, 20, 21]); C2) it would not allow for the development of new physical theories (e.g. [19, 20]; C3) it would contradict free will; C4) it would imply a cosmic conspiracy between (all) physical systems and/or would need fine-tuned initial conditions [e.g. 1, 17, 21]. Proponents of SD can point to the fact that C3) ('No Free Will') and C4) ('Conspiracy') are metaphysically tainted and are therefore not really a physics problem (more on this in Section 5); but C1) ('No Science') and C2) ('No Theories'), if true, would mean the end of SD as a physical solution to Bell's theorem. Recently Hossenfelder and Palmer [11] have proposed their answer to most of these objections, but at least some of their arguments have been found unclear or wanting by others [20-21]. Furthermore, it is clear from the cited articles that researchers can have a quite different understanding of what SD exactly amounts to. To make the situation even more complex, there is a body of literature that focuses on a closely related loophole for Bell's theorem, namely the 'contextuality' loophole [22-25]. Thus, a clarification of the debate would be helpful.

The article is organized as follows. In Section 2 we will define SD, and briefly explain how it could offer a solution to Bell's no-go verdict (we can be brief here because these arguments are well-known). It will prove useful to distinguish two types of SD, which we will term 'strong' and 'soft': they correspond to two very different interpretations of the variables in the defining formula. We will discuss how the two most recent superdeterministic models [8, 12] relate to this classification. A key idea here is that strong SD is only fully described by a hypothetical ultimate ToE (Theory of Everything) that would allow one to predict every event, including human actions, based on the degrees of freedom describing the Planck epoch – surely an elusive theory, yet quite natural to physical thinking. Soft SD corresponds to realistic hidden-variable (HV) theories. In Section 3 we will argue that strong SD can justify soft SD, so that realistic new local hidden-variable theories (HVTs) can be constructed after all, against the 'No Theories' claim. Similarly, we will argue that strong SD can explain 'future input dependence', a concept from [11, 12]. In Section 4 we will present our response to objection C1), the 'No Science' charge, while paying attention to questions raised in [19-21]. Our rationale will hinge, notably, on the idea that detecting statistical dependence is a matter of measurement precision. We will illustrate this idea in a simple model system, namely a collection of colliding hard spheres – a strongly correlated system with an underlying deterministic (and chaotic) dynamics. In Section 5 we will briefly address the problems C3) ('No Free Will') and C4) ('Conspiracy'), and argue that these are, at least in part, metaphysical worries, and that they should be addressed by the relevant philosophical and neuroscientific theories, in tune with [11, 26]. Arguments pro and contra leave us in a Kantian antinomy. In order to tip the balance in favor of SD, we will call probability theory to the rescue in Section 6[1]. The aim of this Section is to show that SD allows to interpret probability, including the probabilistic assumptions at the heart of Bell's theorem, in a unified manner in the classical and quantum realm. This in turn allows us to answer three foundational questions for which the 'no-HVs' orthodoxy remains powerless. Usually one adopts the principle / theory that can answer the most questions. We will end by succinctly commenting on the contextuality-loophole [22-25], and argue that it is again most naturally explained by SD.

Throughout the article we use non-technical language and address issues that have been raised both in the physics and philosophy communities – well aware that this puts us at risk of disappointing members of both. Even if proponents of SD in the physics community ultimately aim at making new theories and new predictions, SD appears to be an eminently interdisciplinary topic, since it invites notions as free will, cause, causality, conspiracy, and the interpretation of probability – and on these philosophy has a much longer tradition, and the most elaborate theories (cf. especially Sections 5 and 6). This article is an attempt at interdisciplinary synthesis, so we count on a modicum of good-will from both sides.

**2. Bell's theorem and superdeterminism: definitions**.

**2.1. General definitions**.

Bell's theorem proves that local HVTs cannot exist – at least not those described by the assumptions (1-3) below made by Bell to derive the Bell inequality. Most people agree that these assumptions are so general that they comprise *all* possible local HVTs – hence the strength of the theorem. But as recalled below, SD rejects one of these premises, for reasons to be explained. 'Local' has a precise meaning in the phrasing of Bell's theorem as we presented it: local HVs belong to local theories, i.e. those that do not involve superluminal influences and comply with Lorentz-invariance and relativity theory. This is the precise meaning we have to give to 'local' in this context: recall the explicit conclusion of

---

[1] Parts of this Section were elaborated in greater detail in [27].

Bell's [28]. For grasping the full strength of Bell's theorem, it is of course important to keep definitions in mind.

In some more detail, the Bell inequality can be derived from the premises (1), (2) and (3) below; let us briefly explain them. In a Bell experiment, one measures correlations $P(x,y|a,b)$, where x and y are the left and right spin or polarization, and a and b the left and right analyzer angles. Bell investigates whether this joint probability can be explained by HV $\lambda$ (having a distribution $P(\lambda)$), i.e., whether one can write equation (1), the 'HV assumption'. (We consider the case of discrete variables, easily generalizable as usual for continuous variables.) Note that (1) follows from elementary probability calculus as soon as one assumes that (x, y) depend on some variables $\lambda$. To derive the Bell inequality, it suffices to further assume the conditions (2) (often called 'locality', or 'factorizability' or 'local causality', which is supposed to follow from relativistic locality) and (3), often called 'measurement independence', 'statistical independence' or 'freedom of choice'. We will use the term 'measurement independence'.

$$P(x,y|a,b) = \sum_\lambda P(x,y|a,b,\lambda)\, P(\lambda|a,b) \quad \text{"Hidden Variables"} \quad (1)$$

$$P(x,y|a,b,\lambda) = P(x|a,\lambda)P(y|b,\lambda) \quad \text{"Locality"} \quad (2)$$

$$P(\lambda|a,b) = P(\lambda) \quad \text{"Measurement Independence"} \quad (3)$$

In order to derive the Bell inequality, we need the average product xy (given the angles a and b) denoted by M(a, b). Only when using all conditions (1-3) can one write (4b):

$$M(a, b) = \sum_{x,y} xy P(x,y|a,b) = \sum_{x,y} xy \sum_\lambda P(x,y|a,b,\lambda)\, P(\lambda|a,b) \quad (4a)$$

$$= \sum_{x,y} xy \sum_\lambda P(x|a,\lambda)\, P(y|b,\lambda) P(\lambda). \quad (4b)$$

But assumptions (1-3) (or (4b)) lead to a contradiction with quantum mechanics and the experiments; *hence at least one of (1-3) must be false*. Note that the deterministic case is included in (1-4): it corresponds to $P(x|a,\lambda)$ given by a Kronecker-delta: $P(x|a,\lambda) = \delta_{x,f(a,\lambda)}$ for a function $f$ (similarly for $P(y|b,\lambda)$).

We will not discuss here all proposed solutions, but the most straightforward one is to reject (1) and adopt Bohr's Copenhagen interpretation, stating that additional variables underneath quantum properties simply do not exist. From a pragmatic point of view, this surely seems a reasonable position. Others, e.g. proponents of Bohmian mechanics, reject (2) and must therefore accept nonlocality (in the strong sense defined above[2]). SD, then, can *broadly* be defined as the position assuming that (3) is not universally valid for HVTs, while locality (2) is satisfied. In other words, if physically sound, according to SD there can exist local HVTs for quantum mechanics after all. In terms of epistemic categories, SD can be understood as a hypothesis (or a principle) and superdeterministic theories / interpretations are theories / interpretations that exploit this assumption.

---
[2] Indeed, a relativistic version of Bohmian mechanics is lacking (see [29], § 14).

SD is, of course, a speculative assumption, for instance because (3) has always been verified for macroscopic systems in comparable Bell-like experiments – confirming most people's intuitions that the λ of any reasonable HV theory are 'free variables'. But, then, the classic interpretations of Bell's theorem leave many people equally unsatisfied. Most importantly, on these classic interpretations Bell's theorem seems a strong obstacle to unifying quantum mechanics and relativity theory, as we will recall in Section 6. A new interpretation that would not be at odds with the unification program would obviously be welcome.

Measurement independence (3) is usually justified by the classic free-will argument: there cannot exist a statistical dependence between experimenters' choices (a, b) and properties λ that determine the outcomes; such a correlation would violate free will. Others consider violation of (3) a blatantly conspiratorial solution (more on this in Section 5). Clearly, these arguments are quite natural, and again, as far as we know (3) is always valid for macro-systems in a comparable Bell-type setting. But even this could be tested, notably in macro-systems that mimic certain quantum properties[3] [30].

**2.2. Superdeterminism, Soft and Strong**.

At this point it is instrumental to note that at least two variants of SD appear in the literature – two ways in which (3) could be violated. To keep things simple, let us refer to them by strong (or Type-I) SD and soft (or Type-II) SD. Strong SD is the classic version that is usually discussed in the literature [1-3] and that locates the origin of the correlation $P(\lambda|a,b)$ before the Bell particles exit their source, possibly at the Big Bang (cf. below). Soft SD comprises all positions that interpret $P(\lambda|a,b)$ in a different way, for instance through 'future input dependence' [11, 12] or through direct interaction λ ←→ (a, b) at the moment of measurement [30]. Note that the symbols (a, b) can a priori denote two physical properties: either the analyzer angles, or the choices of these angles (in experimenters' brains or in random number generators etc.). Strong SD is at ease with both interpretations; soft SD typically interprets (a, b) as the angles themselves. A series of mathematical toy models violating (3) and recovering the quantum correlation for the Bell experiment have been proposed, as reviewed in [31]: these make no commitment to a physical interpretation and could also be classified as soft-superdeterministic. The same holds for 'contextuality' as proposed in [22-25]. One goal of this paper is to disentangle these different types of SD. We will argue that (i) soft SD can be exhibited by *effective* HVT theories, while strong SD could only be calculated within a final Theory of Everything (ToE); and (ii) that strong SD can justify soft SD.

---

[3] Eq. (2-3) are probabilistic assumptions that are surely intuitive, but there is no proof within an accepted physics theory that they necessarily apply to a given physical system.

In defense of a type of soft SD, it has been noted that if λ contains e.g. detector variables, or variables describing a field in the neighbourhood of the detectors, then, at the moment of measurement, λ may well depend on the angles a and b (for instance due to an interaction of the analyzers with these λ) [30]. This is a natural reading if one assumes that the values of all variables in the probabilities in (1-3), including λ and (a, b), obtain at the time of measurement. The problem with this argument is that it is legitimate to apply Bell's reasoning while assuming that the λ take their value when the particles leave the source (this is how Bell envisages the λ in [28]). In that case one expects that (3) (and (1-2)) hold, *at least in time-resolved Bell experiments*. Indeed, in advanced dynamic Bell experiments the angles a and b are activated just before measurement; the particles in the source cannot 'know' which (a, b) they will encounter. In such time-resolved Bell experiments all five variables (x, y, a, b, λ), more precisely the measurements of x and y, the activation of a and b, and the initial values of λ, can all be considered to be mutually spacelike separated, or so it seems (see a detailed exposition of how this is experimentally realized in e.g. [32]). It seems then hard to see how there could be a correlation between the initial values λ and (a, b) – *unless this correlation was initiated before the particles leave the source:* the genuine type of SD. In sum, we believe that the most natural explanation or justification of soft SD is strong SD, a point we aim to consolidate in Section 3.

In [11, 12] the authors argue differently. The superdeterministic model proposed in [12] is a "toy model for an effective limit of a more fundamental theory", and describes a mechanism of decoherence, more precisely of amplitude damping in an N-level system with quantum Hamiltonian H. This is a toy model in that it has no pretense to be a faithful physical description of what happens in quantum systems in general; but it is still a physical model that can teach us a lot. The authors' main goal is to show that in a theory as theirs there is no fine-tuning required to come to the quantum results. The model starts from a dynamical law (of the Lindblad-form) which involves H and an operator L that depends on the hidden variables λ (cf. [12] Eq. (4-6)). It is explicitly (soft-)superdeterministic, since the λ depend on the detector angles, through what is called 'future input dependence': "The key property of the model discussed here is that the evolution of the prepared state depends on what the measurement setting will be at the time of measurement" [12]. It is then shown that this dynamics leads for any kind of quantum system with a finite-dimensional Hilbert space to the same average predictions as quantum mechanics (the Born rule).

Donadi and Hossenfelder acknowledge that there is no explanation given of how the particles can 'know in advance' which detector angles they will encounter – we repeat, a feat particularly mysterious in dynamic Bell experiments. The authors show that future input dependence can gain cogency by studying certain Lagrangian optimization principles, but it appears to rely on an unusual theory of (backward?) causality (cf. [12], §3). We believe this puts it at odds with among the most widely accepted hypotheses of physics and of science in general [27]; as a consequence, it can be feared that valuable investigations as [11, 12] will not meet the reception they deserve. However, in the next Section we will argue that one does not need to invoke future input dependence; it may find a natural explanation in classic, strong SD.

## 3. Can strong SD justify soft SD ? The 'No Theory' argument.

We have just emphasized that it is essential to consider dynamic Bell experiments when speculating about admissible HVTs: such theories must work also for these tests since these have confirmed quantum mechanics. A well-known argument to reject (3) even for such experiments, so an argument for strong SD, starts from noting that the variables (a, b), which can refer to the angles or to the choice events (in brains or randomizers) leading to these angles, represent physical events. In line with modern science, even human choices are neurobiological, and ultimately physical, events in brains. *Now, SD in essence just says that all events are determined by causes* – a classic of philosophy at least since Spinoza. Thus, according to this classic interpretation of SD [3, 26] (at least the one we have in mind), all events are preceded by causes, which have in turn their causes, until the causal tree converges to the Big Bang. In this picture the degrees of freedom describing the Big Bang are causes of all subsequent events. Then, if λ in (3) refers to such 'ab-initio' degrees of freedom (or to other variables in the overlap of the backward light cones of (a, b) and (x, y)), equation (3) does in principle not hold.

Admitted, this is a heavy tale. It seems alas impossible to ever have access to such ab-initio λ functioning as direct causes of (x, y) and (a, b); it seems impossible we can construct an ab-initio ToE involving λ that would allow us to calculate *everything*, also Alice's choice of her detector angle in her next Bell experiment. Note however that it is an utterly natural idea in the physics community that such a ToE, unifying e.g. gravity and quantum mechanics and describing the Big Bang, does exist as a matter of principle. But surely the practical impossibility to calculate $P(\lambda|a,b)$ is – understandably – at the origin of the disliking of SD by physicists. Bell was among the first to express this sentiment: "A theory may appear in which such conspiracies inevitably occur, and these conspiracies may then seem more digestible than the non-localities of other theories. When that theory is announced I will not refuse to listen, either on methodological or other grounds. But I will not myself try to make such a theory" ([33], p. 106). Bell's 'No Theories' verdict seems to have had a lasting impact. But we doubt that this is the end of the story.

The whole question becomes whether the above hard-superdeterministic picture is compatible with a *physical* solution, i.e. one that allows for constructing new – local – theories. In other words, could this (metaphysical) picture be compatible with *'effective'* local HVTs that involve *accessible* variables $\lambda^*$ – effective theories of quantum gravity, effective ToEs unifying all forces, theories such as those envisaged e.g. in [5, 8, 11, 12] ? We suspect that this is indeed the case, and that the mentioned recent investigations actually corroborate this idea [8, 12].

The soft-superdeterministic solution we have in mind seems to lie just in front of us: we just need to acknowledge that the quantum correlation $P(x,y|a,b)$ can also be expanded with variables $\lambda^*$ that are *not* the ab-initio $\lambda$ just mentioned, but that are correlated with (a, b) (and x, y) *through the ab-initio $\lambda$*. Thus the $\lambda$ act as common causes of the $\lambda^*$ and of (a, b, x, y) – which is a manifestation of classic, strong SD. We then have:

$$P(x,y|a,b) = \sum_{\lambda^*} P(x,y|a,b,\lambda^*) P(\lambda^*|a,b), \qquad (5)$$

and

$$P(\lambda^*|a,b) = \sum_{\lambda} P(\lambda^*|a,b,\lambda) P(\lambda|a,b) \neq P(\lambda^*). \qquad (6)$$

This solution is represented as Case 2 in Figure 1, where the $\lambda^*$ are part of an effective HV theory – i.e., a soft-superdeterministic theory.

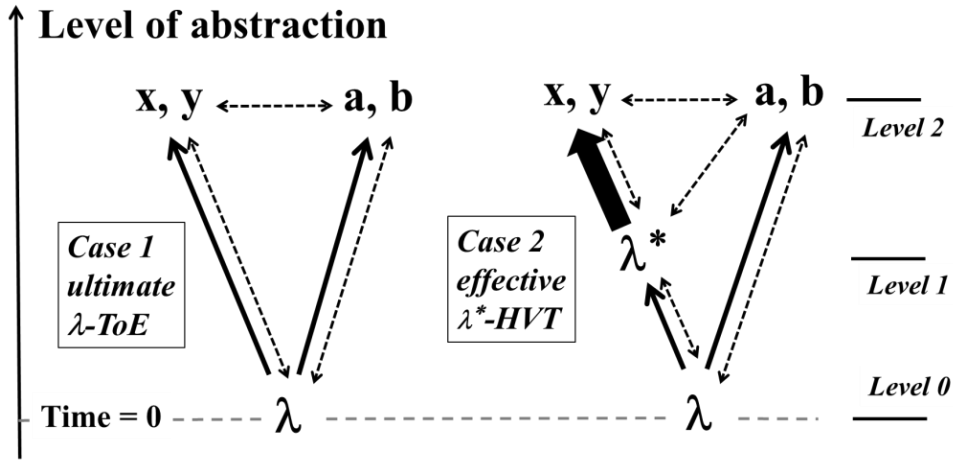

**Figure 1**. Strong SD (Case 1) and soft SD (Case 2), corresponding to two different interpretations of how the superdeterministic correlations $P(\lambda|a,b)$ and $P(\lambda^*|a,b)$ could come about. Case 1 represents the case of an elusive ultimate ToE (involving variables λ), while Case 2 corresponds to a realistic, effective HVT/theory of QG/ToE (involving λ*). Arrows indicate causal connections, and dashed double-headed arrows indicate correlations they generate. The thick arrow symbolizes how an effective $\lambda^*$-HVT could explain quantum mechanics. Variables are part of theories of different 'levels of abstraction' (cf. text).

In this soft-superdeterministic interpretation (a, b) are now typically the detector angles (not the choice events), and (5) is calculable: the quantum properties (x, y) can be explained by hidden variables λ* to which we have access (by theory and/or measurement). Crucially, it seems then that the λ*-HVT can be a *local* theory after all, *since the soft-superdeterministic correlations $P(\lambda^*|a,b)$ allow to trump Bell's no-go verdict*. These correlations $P(\lambda^*|a,b)$ in (5) and (6) exist, even if we have no access to the expansion in (6), since we have no access to the ab-initio λ. Note that the ultimate explanation of why $P(\lambda^*|a,b)$ can be assumed to exist even in dynamic Bell experiments is that these correlations were predetermined by the ab-initio λ. It is in this sense that strong SD still needs to justify soft SD.

In Figure 1 we have associated theories to different 'levels of abstraction': by definition, the ab-initio ToE is of level 0, quantum mechanics of level 2, and the intermediate level 1 is taken by the soft-superdeterministic λ*-HVT. Analogous schemes of reduction between theories of different levels are not rare in physics. For instance, one has such a relation between thermodynamics (a level-2 theory), statistical thermodynamics (level-1), and Newtonian mechanics (a level-0 theory): the variables of thermodynamics can be calculated within statistical physics, which is assuming the existence of the particle variables of mechanics – without needing to know the values of these zillions of variables. (Although not essential for our rationale, note also that there is very well a relationship of (probabilistic) causality between the variables of theories of different levels as indicated in Figure 1. For instance, the fact that the mechanical variables (level-0) take certain values (in some particular case) causes the partition function (level-1) to take a certain value. These causal relations explain where the correlations between levels come from. We emphasize that such a causal interpretation is in agreement with the relevant theories of causality [27].)

Following analogy might be helpful to better understand the legitimacy of soft-superdeterministic theories as represented in Figure 1; we will elaborate aspects of it in the next Section. Assume we live in a purely classical world in which all matter consists of hard-ball atoms as in a classical gas (quantum effects are neglected). Then all macroscopic properties (say x,y), including statistical ones, such as described by statistical mechanics (level-1) or thermodynamics (level-2), and conceivably also all 'freely chosen' analyzer variables (say a, b), are – in principle – correlated with the variables ($\lambda$) of mechanics (level-0) describing the micro-constituents of matter. If we would specify the values of enough properties $\lambda$ of enough micro-constituents, $P(x,y|\lambda)$ and $P(a,b|\lambda)$ would be delta-functions; and more generally $P(x,y|\lambda) \neq P(x,y)$ and $P(a,b|\lambda) \neq P(a,b)$ (cf. next Section). In such a world, correlations of the type $P(a,b|\lambda)$ (and therefore $P(a,b|\lambda^*)$) would exist, even if it may well be that they are unknowable, if the universe is complex enough (we will illustrate this in the next Section). Following this analogy, it is legitimate to assume for such a universe the existence of the correlations $P(a,b|\lambda^*)$ also in a Bell experiment; again they arise through the common causes $\lambda$. And notwithstanding all-pervasive and seemingly conspiratorial correlations among all variables, physics is still possible: level-1 and level-2 theories (statistical mechanics and thermodynamics in the analogy) can still be constructed.

This concludes our argument against the 'No Theories' charge against SD. We believe soft-superdeterministic $\lambda^*$-HVTs can exist: statistical, effective theories that integrate-out the unknowable ab-initio HVs. The link with the toy model of [12] seems still remote, but since this is a model of an "effective limit of a more fundamental theory", where indeed HVs are integrated-out, the link seems not impossible[4]. The same conclusion holds a fortiori for the Cellular Automaton Interpretation of quantum mechanics, the theory to which 't Hooft has devoted efforts since many years [4-8]. This is a particularly ambitious theory, which has the goal to explain quantum mechanics and ultimately to become a Grand Unified Theory, or at least to offer ideas to get there. The theory is not finished but it is clear that highly interesting preliminary results have been obtained. 't Hooft's most recent insights are summarized in [8]. In particular, the author shows there that by positing the existence of fast-fluctuating classical degrees of freedom, possibly shrunk to ultra-small dimensions, and coupled to slow classical degrees of freedom (cf. his Eq. (13) on p. 11), "we can mimic almost any quantum Hamiltonian" [8]. It seems that the link with the scheme in Figure 1 can be made: the role of the ab-initio HVs is taken by the fast variables (which, 't Hooft conjectures, could be the vacuum fluctuations of quantum fields of the Standard Model). What is important for our argument is that it is by integrating-out these variables (cf. [8] Eq. (14) and accompanying text) that quantum mechanics emerges. So again, the theory explicitly posits the existence of strong-superdeterministic variables, evolving via permutations as in a deterministic automaton (cf. [8], Eq. (1)); but we do not have to know, and we cannot know, the details of this gigantic state space; we can integrate them out and construct an effective theory involving new effective HVs (the slow variables). Such an effective theory is an intermediate level between an ultimate ToE and quantum mechanics. Superdeterminism is assumed, and needs to be assumed to counter Bell's no-go argument; and yet effective theories are possible.

---

[4] The model of [12] is in any case a soft-superdeterministic model in the sense that (a, b) refer to angles, not angle choices.

Now, there is a well-known objection to the hypothesis that strong-superdeterministic degrees of freedom cause 'everything' since the Big Bang: in that case everything would be correlated to everything, in tension with the scientific practice. Let us address this problem in the next Section.

**4. The 'No Science' argument. A billiard table as model.**

One of the criticisms of SD states that it would make the usual practice of empirical science impossible. Science is based on statistical inferences and assumes that observed systems and testing procedures can be independent – allegedly in contradiction with the ubiquitous statistical dependence implied by strong SD. This criticism was already phrased by Shimony et al. in 1976 [1]:

> "In any scientific experiment in which two or more variables are supposed to be randomly selected, one can always conjecture that some factor in the overlap of the backwards light cones has controlled the presumably random choices. But, we maintain, skepticism of this sort will essentially dismiss all results of scientific experimentation. Unless we proceed under the assumption that hidden conspiracies of this sort do not occur, we have abandoned in advance the whole enterprise of discovering the laws of nature by experimentation."

Both arguments against [11] and in favor [20, 21] of this criticism have recently been published, but it seems that no consensus has arisen. Let us see if we can clarify the debate.

It is indeed a fact that if one assumes ab-initio common causes for all events, one naturally comes to the conclusion that everything should be correlated with everything else. But we observe statistical *in*dependence between usual variables all the time, of course, and we can demonstrate it with statistical tests. However, this contradiction can be dissolved if one observes 1) that (strong) SD posits the dependence on truly unique variables $\lambda$, namely variables belonging to the ultimate ToE (!), and 2) that *observing / measuring statistical independence is a matter of experimental precision and of the number of causal events involved*. It is well conceivable that Eq. (3) is violated *as a matter of principle* (in an inaccessible ab-initio ToE to which the $\lambda$ belong), while *in practice* one cannot measure the correlations $P(\lambda|a,b)$, just as one cannot measure the statistical dependence between many or most variables (say $P(a|b)$) – even if these correlations exist in theory through the common causes $\lambda$. On the other hand, some correlations *can* be measured, say the quantum correlations $P(x,y|a,b)$ in a Bell experiment. Assuming SD, is there any coherent explanation of when correlations are observable and when not (e.g. for macroscopic variables) ? – a worry raised e.g in [19-21].

If idea 2) is not clear, it can be illustrated in probabilistic model systems by numerical simulation. One such model system is a collection of elastically colliding disks or non-spinning billiard balls, set in motion by an initial collision by a cue ball (Figure 2). The analogy to the real superdeterministic universe is that both systems are set in motion by an initial 'bang'.

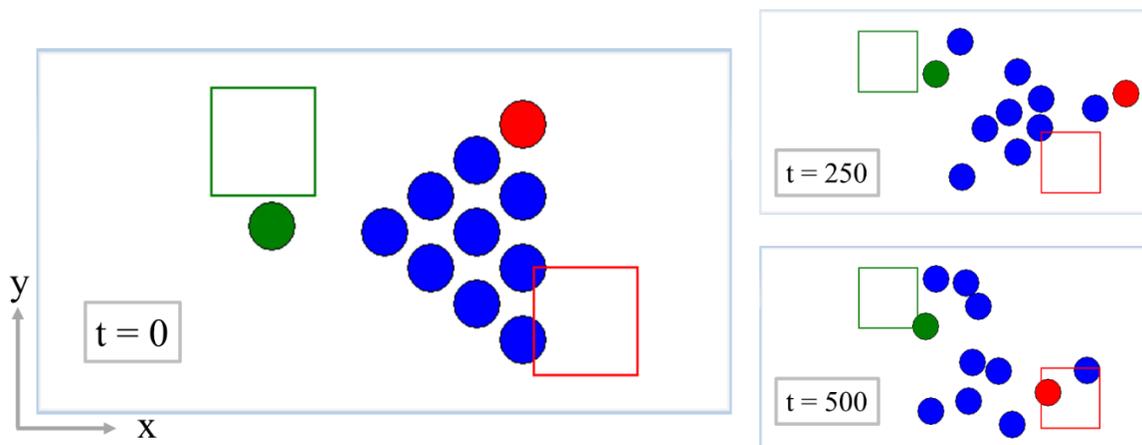

Figure 2. A billiard table. The cue ball (green) hits the other balls at t=0 causing multiple elastic collisions. Calculated positions at t = 250, 500 are also shown.

In vector notation, each 2-ball collision, say between ball 1 and 2, satisfies law (7), based on conservation of momentum, kinetic energy, and angular momentum. Here $v'_{1(2)}$ is the velocity of ball 1(2) after collision, and $v_{1(2)}$ before collision, and $x_{1(2)}$ is the center of ball 1(2) at the moment of collision.

$$v'_1 = v_1 - \frac{\langle v_1 - v_2, x_1 - x_2 \rangle}{\|x_1 - x_2\|^2}(x_1 - x_2)$$
$$v'_2 = v_2 - \frac{\langle v_2 - v_1, x_2 - x_1 \rangle}{\|x_2 - x_1\|^2}(x_2 - x_1). \tag{7}$$

Even if the movement of each ball is perfectly deterministic, one can turn this into a probabilistic system by considering N billiard tables as in Figure 2, and by stochastically varying one or more variables over the tables. Our basic experiment considers a variation of the initial $y_{1,0}$-position of the green cue ball. Since this system is chaotic, slight variations (uniformly distributed over an interval ε) in this variable will, after sufficiently long time, i.e. after sufficiently many collisions, lead to quite different positions and velocities of the balls. In this well-defined probabilistic system, one can calculate – for instance – the probabilities of following events (corresponding to dichotomic variables):

$E_1$ = the event that the center of the green ball has passed at least once through the green test square during the interval (0, t).

$E_2$ = the event that the center of the red ball has passed at least once through the red test square during the interval (0, t).

We calculated the positions and velocities of all balls at every discrete time, so we can numerically determine the following frequencies. These should converge to probabilities for N large enough, as verified below:

$$P(E_i) = \frac{number\ of\ times\ E_i\ occurs\ in\ the\ N\ tables}{N}. \tag{8}$$

Clearly, these probabilities depend on a series of experimental variables, such as the number $N_b$ of balls, the dimensions of the table and the test squares, the interval ε, the initial velocity of the cue ball, the measuring time t, etc. (the parameters we used are given in the Appendix; figures are drawn on scale). We investigated under which conditions there is a measurable correlation between $E_1$ and $E_2$, so when:

$$P(E_1, E_2)\ [= P(E_1|E_2)P(E_2)\ ] \neq P(E_1)P(E_2). \tag{9}$$

The correlation can be quantified by:

$$\Delta = \left| P(E_1, E_2) - P(E_1)P(E_2) \right|. \tag{10}$$

To fix ideas, one can call this the 'correlation between green and red passages'. Of course, in this system a huge number of variables and their correlations can be analyzed; but our results R1)-R3) below are quite generic.

*Note that in this system one expects correlation between all dynamical variables* (taken at a time when enough collisions have occurred), *since these all have common causes*: the 'bang' at t = 0 and the intermediate collisions. Our simulations confirm this, as well as following more or less intuitive results: R1) Observing / measuring statistical (in)dependence is a matter of experimental precision; R2) In order to observe correlation, sufficiently many variables need to be fixed; R3) The correlation $\Delta$ decreases with the number of causal events (collisions) involved.

R1) follows from the fact that the probabilities in (8) and (9) are subject to experimental uncertainty (i.e. experimental, in our case numerical, precision), which decreases with increasing the number of trials, i.e. the number of tables N. Thus, if correlation exists, it can only be shown to exist for sufficiently precise probability measurements. This result is illustrated in Figure 3, which shows the evolution of probability ratios as given by (8) as a function of N (for t = 500). If N is small, the probability ratios do not stabilize and the correlation is not measurable; but eventually the ratios do stabilize – the hallmark of a probabilistic system [34-37]. In this example, we estimate the correlation $\Delta = 0.086 \pm 0.002$.

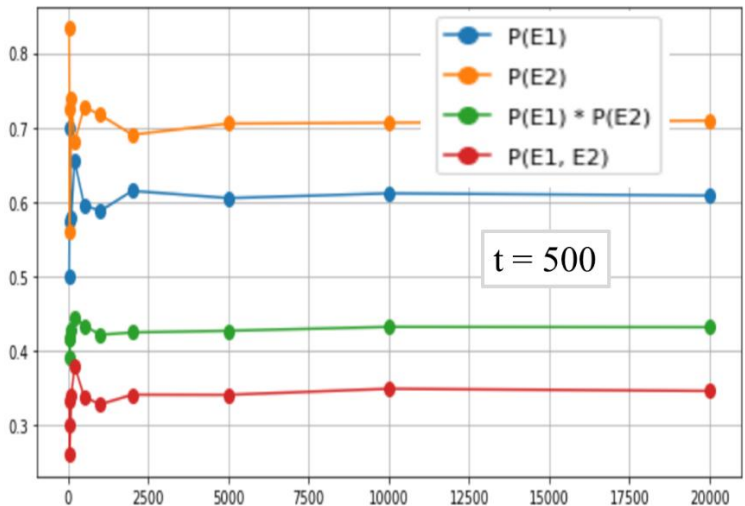

Figure 3. Probability ratios at t = 500 as a function of N
(parameters of the billiard table in Appendix).

This observation R1) becomes clearer if one introduces additional randomness in the system, by assuming a distribution over any additional variables. An example is given in Figure 4, which shows analogous results as Figure 3 for exactly the same conditions except that now the positions and velocities of all balls at t = 0 are uniformly distributed over given intervals; after t = 0 they evolve again according to (7). This corresponds to Brownian motion of the green and red Brownian particles.

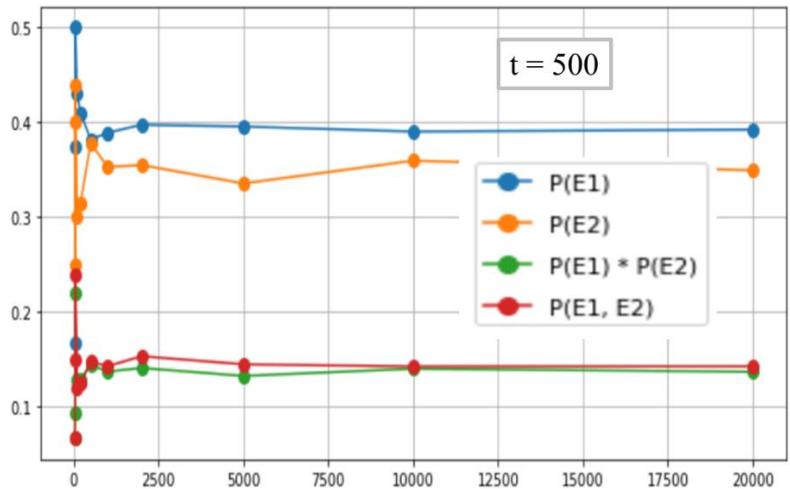

Figure 4. Probability ratios at t = 500 as a function of N
for Brownian motion (same parameters as Figure 3, cf. Appendix).

The correlation $\Delta$ is now hardly visible on the graph; but precise numerical evaluation using up to N = 50,000 trials shows it is non-zero: $\Delta = 0.007 \pm 0.002$. One can add randomness on a series of other parameters, say one of the dimensions of the table, a ball diameter, the time window etc.; in which case the practically measureable correlation goes to zero. Yet it exists *as a matter of principle* due to common causes, i.e. the collisions before measurement. Experiments as these confirm R2): in order to observe correlation, sufficiently many variables need to be fixed.

Finally, R3) is illustrated in Figures 5a and 5b, which show the probabilities using the same parameters as in Figure 3, except that in Figure 5a the measurement time t = 1000 and in Fig. 5b the initial horizontal speed $v_{x,0}$ of the cue ball is multiplied by 5 compared to Figure 3. In Figure 5a $\Delta$ is reduced by a factor of about two compared to Figure 3, and in Figure 5b $\Delta$ practically vanishes. Thus, more collisions tend to wash out correlations.

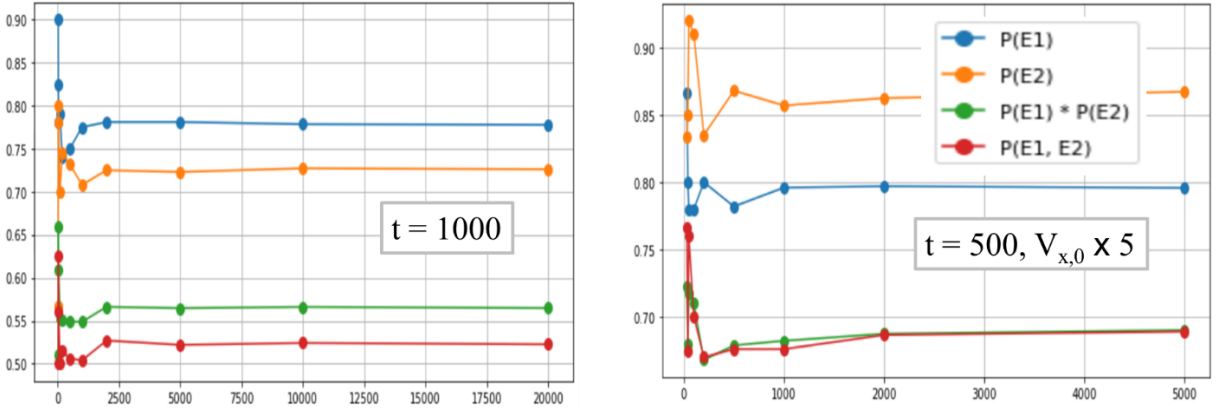

Figure 5a, 5b. Compare to Figure 3. 5a: t = 1000 and 5b: $v_{x,0}$ five times higher.

Let us now finalize the analogy with the real Bang at the beginning of the universe, and return to measurement independence (3). Clearly, statistical independence is ubiquitous in our world, say between Alice's and Bob's 'free' setting choice or between experimental outcomes and the choices of the procedures used to statistically sample these outcomes. But in a superdeterministic picture it is natural to consider uncorrelated or 'free' phenomena as being correlated under the surface, much as in the billiard tables. The correlation is too weak to be noticeable: the common causes are too 'old'; too many causal influences have washed out the effect of the common causes, as illustrated by R3). Similarly, in this picture the superdeterministic correlations $P(\lambda|a,b)$ violating (3) exist as a matter of principle (as they would be revealed by an inaccessible full ToE), but they cannot be measured in practice neither be described by accessible theories. Note that there is another perspective to understand why the quest for correlations between macroscopic variables and sub-quantum variables λ is hard. We know that the Bell correlations $P(x,y|a,b)$ can be measured[5] (just as in the billiard collisions some correlations can be measured). To probe these correlations $P(x,y|a,b)$ between quantum degrees of freedom already demands draconic control over experimental variables. Indeed, as shown in our model calculations, in order to exhibit correlation one needs to fix sufficiently many variables[6] (R2). So probing $P(\lambda|a,b)$ involving sub-quantum variables λ seems even more arduous. It is unknown at the time being which additional variables would have to be controlled in order to exhibit $P(\lambda|a,b)$. But we argued that soft-superdeterministic correlations $P(\lambda^*|a,b)$ could be detectable. In this context the experiment proposed in [9] is essential: here the λ* are hidden detector variables, which might be more easily controllable.

---

[5] A possible reason why one can measure this quantum correlation, in contrast to many of the correlations between macroscopic variables, is that x and y refer to 1-particle variables of a fundamental theory, quantum mechanics. Usual classical systems have too many degrees of freedom and too many interactions between them, so they appear uncorrelated (unless a strong enough cause correlates them). This is a possible answer to questions asked in [19-21].
[6] In quantum experiments decoherence is ubiquitous and one needs to fix not only a and b but a whole series of other variables; a specific well-controlled quantum set-up is needed to detect correlations.

In conclusion, the objection to SD that it would not allow for the usual scientific methodology, is not compelling. Even in a superdeterministic universe, there is no reason to believe that all correlations are detectable, even if they exist.

**5. The conspiracy and 'No Free Will' arguments**.

Let us briefly comment on the well-known objections to SD stating that it implies conspiratorial correlations and that it is contradictory to free will. We will be succinct, since these objections have been countered a number of times recently (e.g. [11, 12, 26]). Moreover, they are philosophically tainted and it seems difficult to come to any consensus. Whether determinism is at odds with free will is an ultra-classic problem of philosophy since centuries and probably millennia. Since about fifty years it also became topical in neuroscience, notably after the seminal experiments by Libet [38]. In these communities three main positions are debated. 'Libertarians' – surely the majority of lay persons in Western countries – reject determinism and believe in an absolute form of free will that is ultimately independent of external causes[7]; 'hard determinists' reject free will as an illusion – we just think we have free will because we cannot grasp the many causes preceding our 'free' choices, indeed zillions of causes according to proponents of strong SD. The latter position is corroborated by remarkable neuroscientific experiments [39]. Although both camps still have many proponents, nowadays a majority of philosophers ([40], p. 242) believe that free will and determinism are *not* contradictory – a position termed 'compatibilism'. Compatibilists have shown that one can meaningfully define free will in a deterministic universe, for instance as a capacity to deliberate about acts and choices in a way that is not constrained by other persons. Compatibilists and determinists typically point to the fact that 'free', indeterministic decisions that pop up in the human mind without any preceding determining factors (factors such as reasons or beliefs), are, by definition, purely random and therefore erratic. Is this rational free will? Is this free will that anyone wants? Determinists and compatibilists prefer to see decisions as embedded in neural circuits that are themselves connected to, and under causal influence of, neural nets that materialize for instance reasons, beliefs or external inputs.

---

[7] Such a view immediately begs the question: what is the entity that materializes, or is responsible for, such an absolutely independent free will? An immaterial mind? A homunculus in the head? There is no evidence whatsoever for such a spiritual entity in neuroscience.

We have given an analysis of these metaphysical questions, compatible with neuroscientific findings and physics, in [41, 42]; we did not find any reason to consider (super)determinism as antagonistic to free will, rather to the contrary. Therefore we believe that considering Eq. (3) as imposed by free will is not in tune with specialized research.

Now the charge of conspiracy, a tougher problem. First note that this is an objection against *strong* SD. Indeed, in [12] it has been shown that soft-superdeterministic theories cannot be accused of this problem, neither of the closely related fine-tuning. Regarding strong-SD, we admit: the idea that the ab-initio HVs are correlated with (a, b) in precisely such a way that the expansion (1) always leads to the quantum result, whatever the process by which the choices (a, b) are determined (by humans or sophisticated protocols of randomization), surely *looks* mind-boggling. At least from one perspective. But five centuries ago, theories as quantum mechanics, the neural conception of the brain, and the mechanics of moon travel would have looked mind-boggling too to our ancestors. And another perspective, based on the assumption that the universe is a deterministic system starting from a singularity, is logically possible. Indeed, if such a universe is compatible with the emergence of humans and indeed scientists – perhaps a stark assumption but not a mind-boggling one –, and if at some point in time some correlations as $P(x,y|a,b)$ still exist in this universe even a long time after the singularity took place (both in the billiard system and in our universe many do), and if scientists can detect these correlations, then one must simply expect that they are invariant under experimental details, i.e. how precisely – by which process – the key experimental parameters (a, b) come about. If not, science would not exist. Ergo, if our universe is deterministic, $P(x,y|a,b)$ must be invariant under experimental details of preparation, there is nothing mind-boggling about it.

In sum, even if we admit that our argument is not final, it seems to us that conspiracy is in the eye of the beholder. Let us conclude thus: the verdict depends on the (metaphysical) assumptions from which one starts; the matter is as metaphysical as the Kantian antinomies.

And yet, we believe there is one theory that can shed a more decisive light on the matter, showing that SD is the more promising assumption, compared to the Copenhagen interpretation with its ban on HVs, and compared to nonlocal theories with their well-known tension with relativity. This theory is probability theory, as we have argued in detail elsewhere [27]. For reasons of completeness, we will succinctly summarize the argument in the next Section.

## 6. The final argument: the interpretation of probability.

Probability theory can be understood not only as a mathematical but also as a physical theory: it can be applied – as zillions of tests have shown – to describe physical experiments. The physical interpretation of probability has been elaborated in most detail by von Mises [34, 35] (Kolmogorov refers to von Mises as the primary source for the physical interpretation of probability in his standard work [43]). Now, following idea seems largely absent from the literature: as a physical theory probability theory is, in a sense, a unifying theory, since the outcomes of experiments of both quantum mechanics and relativity theory have to comply with it[8] (a deterministic theory as relativity is a special case of a probabilistic theory, with probabilities 1 or 0) [27]. From this point of view, probability theory is arguably the most encompassing physical theory we have. That is the main reason why, if probability would point to SD, this indication should be taken utterly seriously. There are actually two indications.

First, let us emphasize that measurement dependence (violation of (3)) is compatible with a precise application of probability theory (as a physical theory), while measurement independence is not. Indeed, in recent studies, following von Mises, it has been stressed that 1) probabilities belong to experiments and not to objects or events per se, and 2) that any probability depends *at least in principle* on the 'context' including all detector settings of the probabilistic experiment [27, 36, 37]. Remarkably, it suffices to clearly define what a probabilistic system is, to come to this conclusion. Bohr believed this for quantum systems, but it arguably holds also for classical probabilistic systems. According to this analysis, then, $P(\lambda)$ in (3) and (4b) should be replaced by $P(\lambda|a,b)$ as in (4a): each detector combination (a, b), (a', b) etc. corresponds to a different experiment, hence to a different physical system. (This seems to be in essence the same rationale, also based on the interpretation of probability but rather inspired by Kolmogorov, of authors who point to the 'contextuality loophole' to counter Bell's no-go verdict, see e.g. [22-25].) *Indeed, note that it is (4a) that immediately follows from probability theory, not (4b),* through the rule of total probability. This is not an innocuous implication if one takes probability as a fundamental physical theory.

Now, as with soft-SD, the problem with this explanation is that it is difficult to accept that such a correlation $P(\lambda|a,b)$ comes about in dynamical experiments with a spacelike separation between the detector settings' choices (a, b) and the particles at the source ($\lambda$). But there exists an explanation that is fully compatible with physics, namely strong SD: in a slogan, these correlations already existed. In sum: one can de facto rely on probability theory to reject or at least doubt that (3) and (4b) are correct (since they are certainly not an immediate consequence of the calculus and physical interpretation of probability); and the ultimate justification of this rejection is given by strong SD. One might object that this is 'just metaphysics', since SD, contextuality, future input dependence are all just different interpretations of the same mathematics, namely the rejection of (3). True, but as the reader has understood, this article takes it that metaphysics and physics go hand in hand, and that the former might guide the latter (and vice versa !). SD has already led to proposals for experiments [9].

---

[8] This implies for instance that two statistically independent stochastic variables x and y with an underlying deterministic (relativistic) dynamics must be functions $x = f_x(\lambda_x)$, $y = f_y(\lambda_y)$, where the sets of variables $\lambda_x$ and $\lambda_y$ share no variables.

Indeed, the next indication is the following: *there are at least three questions from the interpretation of probability theory that can only be answered coherently if one assumes SD* – at least if we bar nonlocal HVs. Nonlocal HVs can also answer the three questions below, but there is a heavy suspicion that they are unphysical.

The first question is: *How to interpret probability and statistical correlation in a coherent manner?* – understood: ideally in a way valid for quantum and classical systems. In quantum field theory correlation can be explained by interaction, but for the emergence of probability no further explanation is given: probability is not reduced to something more fundamental. For classical systems, there is a whole body of literature that corroborates the view that determinism explains (classical) correlation, and, to start with, (classical) probability. Indeed, this assumption is actually a ubiquitous guidance principle in the statistical theory of causal modeling [44, 45]. On this view, probabilities as P(x) can ultimately be reduced to functions from $\lambda$ to x, where the set of variables $\lambda$ contains unobserved variables; the $\lambda$ are termed the causes of x ([44], p. 26). The usual interpretation of correlation is that probabilistic dependence between variables x and y is due to common, possibly hidden, causes: either x(y) causes y(x), or these variables share common causes $\lambda$ (or a combination of both). Moreover, a full set $\lambda$ of common causes can de-correlate x and y: $P(x,y|\lambda) = (x|\lambda)P(y|\lambda)$, often called Reichenbach's principle [44, 45]. This classic assumption goes into the condition (2) of Bell's analysis. This causal interpretation of correlation appears to always work for classical systems, and of course it also works for the billiard system of Section 4 (see examples there). For instance, it is easy to verify in this system that if one fixes all variables, the ubiquitous correlation vanishes. Now, this interpretation is believed not to work for quantum correlations, precisely because of Bell's analysis. But if one assumes SD Bell's analysis is trumped and the same explanation of correlation can be upheld for quantum systems. Clearly, the 'no-HVs' assumption of quantum mechanics does not offer a coherent explanation of probability and correlation.

The second question is: *How can one explain the ubiquitous occurrence of the Gaussian probability distribution in classical and quantum systems?* The Gaussian distribution is ubiquitous also in quantum systems [27]. The answer is well-known from applied probability and the Central Limit Theorem, sometimes considered the 'unofficial sovereign of probability theory' [46]. This theorem states that (under broad conditions, notably the Lindeberg condition) any random variable that is the average, i.e. the weighted sum, of many independent random variables, has a normal distribution independently of the distribution of the contributing variables. So the answer to our question is immediate for classical systems: many variables are the average, i.e. a (sum) function, of many other variables, hidden or not; in other words, they are caused by many (hidden) variables λ; and therefore they are normally distributed by the Central Limit Theorem. On the Copenhagen orthodoxy this explanation does not work for quantum Gaussians, but that is again only if one dismisses SD. SD allows one to offer the same explanation in the quantum and classical domain. Just as for question 1, superdeterministic HVs, or hidden causes if one prefers, provide the explanation.

The third question is closely related to Bell's theorem and to one of the greatest current challenges of physics: *Are there degrees of freedom that could unify (in the usual sense) quantum field theories (QFT) and general relativity (GR), and if so, can we specify them (even a little more, qualitatively)?* Of course, our answer will necessarily be speculative[9]. But in the present context it is relevant that an answer can be given based on *bona fide* physics, and coherent with the interpretation of quantum mechanics and probability discussed in this article. Unification of two theories in the usual sense means that the master equations (laws) of both theories can be derived, as special cases, from more general equations, so that the variables of these two theories can be reduced to the degrees of freedom of the unified theory. In Section 3 we discussed this type of reduction (cf. Figure 1). We also have such a reduction of the variables of say optics and magnetism, which all can be expressed in function of the degrees of freedom of Maxwell's electromagnetism. If unification in this sense is possible for QFT, the unified theory (a ToE) should be a HVT, by the above definition of unification. Most likely it should also be a local HVT, since it needs to give rise to GR that is incompatible with nonlocal, i.e. superluminal influences. Ergo, by Bell's theorem, the natural conclusion is that the degrees of freedom that could unify QFT and GR must be part of a superdeterministic theory. It is widely assumed that such a ToE should describe the Big Bang or instants closely after, the Planck epoch. Note that its variables largely qualify as lying in the overlap of the backward light cones of any Bell experiment, and therefore as variables that can, in principle, violate Eq. (3) and thus provide a solution for Bell's no-go verdict. To the contrary, the 'no-HVs' assumption seems to leave Bell's theorem rather as an obstacle to the unification of GR and QFT.

---

[9] If one finds our answer too speculative, note that to come to our final conclusion it is enough that the first two questions are answered.

In conclusion, three foundational questions can only be answered by the assumption of superdeterministic HVs; the no-HVs assumption remains powerless to answer them. (If a superdeterministic HVT for quantum mechanics can be built, as we assume[10], then any questions that quantum mechanics can answer can be answered by the superdeterministic HVT.) True, the three questions and answers are not fully part of physics proper. Nevertheless, also in philosophy of physics one usually adopts those principles and theories (if well-embedded in physics) that can answer the most questions. We believe this puts SD ahead of its competitors, at least as a guiding principle for building new theories.

## 7. Conclusion.

The aim of this article was to make some aspects of superdeterminism more intuitive, and to add some clarification to debates in the literature. We introduced the distinction between soft and strong superdeterminism: two very different conceptions of how to understand measurement dependence, i.e. the violation of Eq. (3), the defining feature of superdeterminism. We argued that soft-superdeterministic HVTs are possible as effective theories involving effective degrees of freedom $\lambda^*$. To justify that such effective HVTs can be constructed, we assumed the existence of an ab-initio ToE with variables $\lambda$, arguably strongly-superdeterministic, which relate to the $\lambda^*$ via $P(\lambda^*|a,b) = \sum_\lambda P(\lambda^*|a,b,\lambda) P(\lambda|a,b)$ (Eq. (6)). Others have invoked 'future input dependence' to explain measurement dependence [11, 12], but this may be an unnecessary move if (strong) superdeterminism suffices as an explanation. Next, we argued that the No Science objection against superdeterminism can be rebutted by noting that in a deterministic universe statistical dependence can exist as a matter of principle, but be undetectable for all practical purposes. We illustrated this claim in a probabilistic model system with deterministic dynamics, namely a billiard game. Further we succinctly presented our view on the No Free Will and Conspiracy objections – emphasizing that these are metaphysically loaded, and that therefore it is necessary to study them through the lens of cogent theories of philosophy and neuroscience. Finally, we argued that the interpretation of probability theory privileges superdeterminism over its main competitors.

---

[10] Based, notably, on the proof given in [3].

**Appendix. Numerical data for the billiard simulations**.

Here we provide the numerical data for the billiard simulations presented in Section 4. Figure 2 is an on-scale drawing of the billiard table at simulation time t = 0, containing 11 balls at rest with radius 20 (all parameters are in arbitrary units). Other parameters of the experiment:

- Dimensions of the table: 600 x 300
- Length of a time tick: 0.1 (experiments can run till t = 250, 500, 1000 etc., cf. text)
- Initial positions of the balls (given as (x, y) coordinates of the balls' centers assuming that (0, 0) is the lower left corner of the table):
  - green cue ball: (200, 155 ± ε), ε = 3 (uniform distribution, cf. text)
  - other balls: (300, 150), (340, 120), (340, 180), (380, 210), (380, 150), (380, 90), (420, 240) [red ball], (420, 180), (420, 120), (420, 60)
- Initial velocity of the green ball: $(v_x, v_y) = (20, 0)$
- Positions of the test squares (given as $(x_1, y_1)$ – upper left corner, and $(x_2, y_2)$ – lower right corner):
  - (150, 270), (240, 180)
  - (430, 120), (520, 30)

For the experiment of Figure 4 (Brownian motion), all the positions of all balls are uniformly distributed with x ∈ [20, 580] and y ∈ [20, 280] (intersections of the balls are not allowed). The initial velocities are distributed uniformly such that $v_x$ ∈ [-20, 20] and $v_y$ ∈ [-20, 20].

**References**.